\newcommand{\al}{\alpha} 
\newcommand{\Ga}{\Gamma}
\newcommand{\ep}{\epsilon} 
\newcommand{\de}{\delta}\newcommand{\aw}{x^i_0+\dot x^i_0}
\newcommand{\vphi}{\varphi}
\newcommand{\R}{{\Bbb R}}
\newcommand{\pa}{\partial}
 
\newcommand{\DD}{{\cal D}}

\newcommand{\fr}{\frac{1}}\newcommand{\il}{\int\limits}
\newcommand{\nn}{\nonumber}
\newcommand{\ba}{\begin{array}}\newcommand{\ea}{\end{array}}
\newcommand{\beq}{ \begin{equation} }\newcommand{\eeq}{ \end{equation} }
\newcommand{\bea}{\begin{eqnarray}}\newcommand{\eea}{\end{eqnarray}}
\newcommand{\beqn}{ \begin{equation*} }
\newcommand{\lb}{\label}\newcommand{\rf}{\ref}

\newtheorem{prf}{{\it Proof}}

\documentstyle[a4,aps,amsfonts,prb]{revtex}
\title{\begin{flushleft} 
       {\small{\bf J. Math. Phys., Vol 39, 1998}\quad
       Report No: UWThPh-1997-38\quad ESI-Preprint No: 505\newline}
        \end{flushleft}
\qquad\\
\qquad\\ 
       Geodesics and geodesic deviation for impulsive\\ gravitational waves}
\author{R. Steinbauer}
\address{Institute for Theoretical Physics, University of Vienna\\   
        Boltzmanng. 5, A-1090 Wien, Austria\\
        E-mail: stein@doppler.thp.univie.ac.at
        }
\begin{document}
\maketitle
\thispagestyle{empty}

%%%%%%%%%%%%%%%%%%%%%%%%%%%%%%%%%%%%%%%%%%%%%%%%%%%%%%%%%%%%%%%%%%%%%%%%%%%%
%abstract           %%%%%%%%%%%%%%%%%%%%%%%%%%%%%%%%%%%%%%%%%%%%%%%%%%%%%%%%
%%%%%%%%%%%%%%%%%%%%%%%%%%%%%%%%%%%%%%%%%%%%%%%%%%%%%%%%%%%%%%%%%%%%%%%%%%%%

\begin{abstract}
\noindent
{\footnotesize
(Received 4 November 1997; accepted for publication 19 November 1997)}
\vskip10pt
\noindent    
The geometry of impulsive pp-waves is explored via the analysis of the geodesic
and geodesic deviation equation using the distributional form of the metric.
The geodesic equation involves formally ill-defined products of distributions 
due to the nonlinearity of the equations and the presence of the 
Dirac $\de$-distribution in the space time metric. Thus, strictly speaking, 
it cannot be treated within Schwartz's linear theory of distributions. To cope 
with this difficulty we proceed by first regularizing the $\de$-singularity,
then solving the regularized equation within classical smooth functions and,
finally, obtaining a distributional limit as solution to the original problem.
Furthermore it is shown that this limit is independent of the 
regularization without requiring any addi\-tional condition, thereby 
confirming earlier results in a mathematical rigorous fashion. 
We also treat the Jacobi equation which, despite being linear in the deviation
vector field, involves even more delicate singular expressions, like the 
``square'' of the Dirac $\de$-distribution. Again the same regularization 
procedure provides us with a perfectly well behaved 
smooth regularization and a regularization-independent distributional limit.
Hence it is concluded that the geometry of impulsive pp-waves can be described
consistently using distributions as long as careful regularization procedures
are used to handle the ill-defined products.
PACS-numbers:\,04.20.Cv,\,04.20.-q,\,02.20.Hq,\,04.30.-w
\end{abstract} 

%%%%%%%%%%%%%%%%%%%%%%%%%%%%%%%%%%%%%%%%%%%%%%%%%%%%%%%%%%%%%%%%%%%%%%%%%%%%%
%first section                     %%%%%%%%%%%%%%%%%%%%%%%%%%%%%%%%%%%%%%%%%%
%%%%%%%%%%%%%%%%%%%%%%%%%%%%%%%%%%%%%%%%%%%%%%%%%%%%%%%%%%%%%%%%%%%%%%%%%%%%%
\section{Introduction}

Plane fronted gravitational waves with parallel rays (pp-waves) are
spacetimes charac\-teri\-zed by the existence 
of a covariantly constant null vector field, 
which can be used to write the metric tensor in the form~\cite{ehlers+kundt}
\beq ds^2\,=\,H(u,x,y)du^2-du\,dv+dx^2+dy^2,\eeq
where $u,\,v$ is a pair of null coordinates ($u=t-z$,\, $v=t+z$) 
and $x,\,y$ are transverse (Cartesian) coordinates.  
In this paper we shall deal especially with impulsive (in the diction of 
Penrose~\cite{penrose}) pp-waves where the profile function $H$ is proportional
to a $\de$-distribution, i.e., takes the form 
$H(u,x,y)\,=\,f(x,y)\,\de(u)$,
where we leave the (smooth) function $f$ of the transverse coordinates 
arbitrary for the moment. This metric is flat everywhere exept on the null
hypersurface $u=0$, where it has a $\de$-shaped ``shock''. 

Such spacetimes arise most prominently as ultrarelativistic
limits of black hole geometries as first derived by Aichelburg and 
Sexl for the Schwarzschild case~\cite{as,bn,ls}. 
On the other hand Penrose has given a more intrinsic
description of such spacetimes by his ``scissor and paste''
approach~\cite{penrose1}, which essentially consists of glueing together two
pieces of Minkowski spacetime along the null hypersurface $u=0$ with a shift in
the $v$-direction. A similar idea was used by Dray and t'Hooft\cite{dt}, 
who introduced a coordinate shift along geodesics in Minkowski space time 
to rederive the Aichelburg-Sexl geometry as well as Penrose's junction 
conditions from the field equations.

The philosophy of the present work is somehow complementary. We take the 
$\de$-shaped metric literally and try to explore the properties of this 
geometry via investigation of geodesics and the geodesic deviation, thereby
following the approaches of Ferrari, Pendenza and Veneziano~\cite{fvp} and
Balasin~\cite{ba}. 

The main purpose of this work is to deal with the singular, i.e., 
distributional quantities in a mathematically rigorous fashion. A thorough
analysis of the geodesic equation shows that it involves ill-defined
products of the $\de$-distribution with the step function.
This difficulty can be circumvented by a proper regularization procedure 
providing us with a perfectly well behaved smooth approximation, which has a 
distributional limit coinciding with the earlier 
results~\cite{fvp,ba}. However, we neither have to 
impose ``multiplication rules'' like $\de\theta=(1/2)\,\de$, nor to use any
additional requirements as the constancy of the norm of the geodesic's tangent
vector across the shock. In fact, this property comes out as a result in our
approach. The details of this calculation are given in Sec.~\ref{g}.

Further investigations show that the Jacobi equation involves ill-defined terms
of an even worse type, such as the ``square'' of the Dirac $\de$-distribution 
and the square of the step function times the $\de$-distribution. However, our 
regularization strategy again provides us with a smooth approximation 
and a well behaved distributional limit, even in this case, where it 
seems to be hopeless 
to use ``ad-hoc extensions'' of Schwartz's linear distribution theory such as 
special ``multiplication rules''. We give the details of the rather lengthy
calculation in Sec.~\ref{j}. Our  distributional ``solution'' of the 
Jacobi equation fits perfectly well in the (heuristically) expected picture, 
showing the consistency of our approach. Finally, we make some comments 
and give an outlook to future work in Sec.~\ref{c}.
%%%%%%%%%%%%%%%%%%%%%%%%%%%%%%%%%%%%%%%%%%%%%%%%%%%%%%%%%%%%%%%%%%%%%%%%%
\section{Geodesic Equation}\lb{g}             %%%%%%%%%%%%%%%%%%%%%%%%%%%%%
%%%%%%%%%%%%%%%%%%%%%%%%%%%%%%%%%%%%%%%%%%%%%%%%%%%%%%%%%%%%%%%%%%%%%%%%%%%
We start with an impulsive pp-wave metric of  the form 
\beq \label{metric}ds^2\,=\,f(x^i)\,\de(u)\,du^2-du\,dv+(dx^i)^2\,\,, \eeq
where $x^i$ ($i=1,2$) denote the transverse coordinates. It is 
straightforward to derive the geodesic equation. 
The nonvanishing Christoffel symbols are
\beq \Ga^v_{uu}=\,-f\,\,\dot\de\,,\quad
     \Ga^i_{uu}=-\frac{1}{2}\,\pa_if\,\,\de\,,\quad  
         \Ga^v_{ui}\,=\,\Ga^v_{iu}\,=\,-\pa_if\,\,\de\,,\,\,
     %\Ga^z_{uu}\,=\,-\frac{1}{2}\,\frac{\pa f}{\pa z}\,\de \,,\,\, 
     %    &\,&\Ga^v_{uz}\,=\,\Ga^v_{zu}\,=\,-\frac{\pa f}{\pa z}\,\de \,,\,\,         
\eeq           
where we have denoted the partial derivatives of $f$ by $\pa_if$ and the
derivative of the $\de$-distribution by $\dot\de$.
Hence we get the equations
\bea u''&=&0\,,\nn\\
     v''&=&f\,\dot \de+2\left(\pa_i\,f\,x^i\,'
                %+\frac{\pa f}{\pa z}\,\dot z
                \right)u'\,\de\,,\nn\\
     x^i\,''&=&\frac{1}{2}\pa_i\,f\,u'\,^2\de\,,\,\,
%    z=\frac{1}{2}\frac{\pa f}{\pa z}u^2\de\,\,,
\eea
where $\,'$ denotes the derivative with respect to an affine parameter and
summation over $i$ is understood. 
We use the first equation to introduce $u$ as a new affine parameter (there
excluding trivial geodesics parallel to the shock hypersurface) 
to get
\bea\label{geo}
     \ddot v(u)&=&f(x^i(u))\,\dot\de(u)
              \,+\,2\,\left(\pa_i\,f(x^i(u))\,\,\dot x^i\,(u)
                         % \frac{\pa}{\pa z}\,f(y(u),z(u))\,z'(u)
              \right)\,\de(u)\,\,,
             \nn \\
     \ddot x^i\,(u)&=&\frac{1}{2}\,\pa_i\,f(x^i(u))\,\,\de(u)\,\,,
%     z''(u)&=&\frac{1}{2}\,\frac{\pa}{\pa z}\,f(y(u),z(u))\,\de(u)\,\,,
\eea
where $\dot{}$ again denotes the derivative with respect to $u$ and we have 
inserted all the dependences explicitly.
Equations~(\ref{geo}) form a system of three coupled, nonlinear ODEs of second 
order in the vectorspace $\DD'$ of distributions. 
For $u\not=0$ all the right hand sides vanish, which is clear from the 
form of the metric tensor, and we expect the geodesics to be broken,
possibly refracted straight lines. 
However, if we take a closer look at system~(\rf{geo}) we see immediately that
the first equation cannot be taken literally in the sense of distributions 
as the terms $\dot x^i\,\de$ involve ill-defined products of the
$\de$-distribution with the step function. 

To analyze the situation in some more detail we integrate
the last two equations using the
(distributional) identity $f(x^i(u))\,\de(u)=f(x^i(0))\,\de(u)$ %to get 
%~\footnotemark[1]
%\footnotetext
%{\vbox{
to get
\beq~\lb{hoppala}
     x^i(u)\,=\mbox{ initial values }+\frac{1}{2}\,\pa_i\,f(x^i(0))\,u_+\,\,,
\eeq
where we have denoted the ``kink''-function $u\,\theta(u)$ by $u_+$.
Note, however, that distributions can only be multiplied by 
$C^\infty$-functions, whereas it's not clear a priori that 
the solution $x^i(u)$ will be smooth; in fact as suggested by 
equation~(\ref{hoppala}) and as shown later in our calculations the solution
will not even be differentiable at $u=0$.
If we still try to go on by brute force it comes even worse: 
inserting~(\ref{hoppala}) into the first equation~(\rf{geo}) we see that the
term $\dot x^i(u)\,\de(u)$ gives rise to the ill-defined product
$\theta\de.$ (For some further comments on this product see~\cite{mo}, p 21.)
\vskip12pt
%%%%%%%%%%%%%%%%%%%%%%%%%%%%%%%%%%%%%%%%%%%%%%%%%%%%%%%
%%% regularization %%%%%%%%%%%%%%%%%%%%%%%%%%%%%%%%%%%%
%%%%%%%%%%%%%%%%%%%%%%%%%%%%%%%%%%%%%%%%%%%%%%%%%%%%%%%
To overcome the undefinedness described in detail above, we apply a careful
regularization procedure. More precisely we regularize the $\de$-distribution 
by a (standard) mollifier or model $\de$-net, i.e., a net $\rho_\ep$,
defined as follows. Let $\rho$ be a smooth function with support contained in
the interval $[-1,1]$ and $\int\rho=1$; now put $\rho_\ep(x)=(1/\ep)\,
\rho(x/\ep)$, $(\ep>0)$. 
Hence system~(\ref{geo}) takes the regularized form
\bea\label{georeg}
     \ddot v_\ep(u)&=&f(x^i_\ep(u))\,\dot\rho_\ep(u)
              \,+\,2\,\pa_i\,f(x^i_\ep(u))\,\,\dot x^i_\ep(u)
                         % \frac{\pa}{\pa z}\,f(y(u),z(u))\,z'(u)
              \,\rho_\ep(u)\,\,,
             \nn \\
     \ddot x^i_\ep(u)&=&\frac{1}{2}\,\pa_i\,f(x^i_\ep(u))\,\rho_\ep(u)\,\,.
%     z''(u)&=&\frac{1}{2}\,\frac{\pa}{\pa z}\,f(y(u),z(u))\,\de(u)\,.
\eea
Next we choose initial conditions in $u=-1$ (i.e. ``long before the shock''), 
more precisely
\bea\label{icreg}
     v_\ep(-1)\,=\,v_0, &\quad&  x^i_\ep(-1)\,=\,x^i_0\,\,,\nn\\
     \dot v_\ep(-1)\,=\,\dot v_0, &\quad& \dot x^i_\ep(-1)\,=\,\dot x^i_0\,\,,
\eea
for all $\ep$. Note that choosing initial conditions in $u=0$ would mean to
``start at the shock'' and one cannot expect to (and indeed does) end up with 
a regularization independent result in this case. 

Due to the regularization procedure we now have to deal with the fully
nonlinear character of the last equation of~(\ref{georeg}). However, as shown 
in appendix~\ref{a1} the special form of the right hand side of the
equation guarantees global existence of the solutions for small $\ep$.
These are given implicitly by 
\bea\label{solreg}
     v_\ep(u)&=&v_0+\dot v_0\,(1+u)+(\theta*\theta*\ddot v_\ep)\,(u)\,\,,\nn\\
     x^i_\ep(u)&=&x^i_0+\dot x^i_0\,(1+u)+(\theta*\theta*\ddot x^i_\ep)\,(u)
     \,\,,
\eea     
where ``$*$'' denotes convolution.

We are going to calculate the (distributional) limits of the 
solutions~(\ref{solreg}) as $\ep$ tends to zero. 
Since distributions supported in an acute cone form a convolution algebra
(where, in particular, convolution is a separately countinuous operation)
%from $\E'\times\DD'\to\DD'$ and (??????????)
it suffices to calculate 
the limits of the right hand sides of~(\ref{georeg});
the distributional limits of the solutions~(\ref{solreg}) are then computed
simply by integration.
\vskip12pt

We begin with the latter two equations of system~(\ref{georeg})
and choose a test function $\varphi$. We have to calculate the limit of
\beq \frac{1}{2}\,\int\limits_\R\,\pa_i\,f(x^i_\ep(u))\,\rho_\ep(u)\,\vphi(u)
     \,du\,
    =\,\frac{1}{2}\,\int\limits_{-1}^1\pa_i\,f(x^i_\ep(\ep u))\,\rho(u)\,
    \vphi(\ep u)\,du\,\,.
\eeq    
Since (for small $\ep$) $x_\ep(u)$ is bounded uniformly on compact sets
%$f$ is smooth, its derivatives are
%bounded in a neighbourhood of $x^i_\ep(0)$ and 
(see Appendix~\ref{a1})
we can use Lebesgue's dominated convergence. Hence the only term we have 
to compute is
\bea\lb{eq11}
    \lim_{\ep\to 0}x^i_\ep(\ep u)&=&\lim\Big(x^i_0+\dot x^i_0\,(1+\ep u)
       +\frac{1}{2}\,\int\limits_{-\ep}^{\ep u}\,\int\limits_{-\ep}^v
       \pa_i\,f(x^i_\ep(s))\,\rho_\ep(s)\,ds\,dv\,\Big)\nn\\
     &=&x^i_0+\dot x^i_0\,+\frac{1}{2}\,\lim\,\underbrace{\int\limits_
       {-\ep} 
       ^{\ep u}\,\int\limits_{-1}^{v/\ep}\,\pa_if(x^i_\ep(\ep s))\,\rho(s)\,
       ds\,dv \,\,.}
     _{\mid\int\int\mid\,\leq\,\|\pa_if\|\,\|\rho\|\,\int\int\,ds\,dv}
\eea
For $\ep\to 0$ the last integral gives zero, since in the limit the range of 
integration only covers a set of zero measure. Hence we get (as expected)
\beq \ddot x^i_\ep(u)\,=\,\frac{1}{2}\,\pa_i\,f(x^i_\ep(u))\rho_\ep(u)\,\to\,
     \frac{1}{2}\,\pa_i\,f(\aw)\,\de(u)\,\,,
\eeq     
within distributions.

Next we turn to the ``critical'' first equation in~(\ref{georeg}) and again
calculate the limit of the right hand side, i.e., of the expression
\bea \ddot v_\ep(u)&=&f(x^i_\ep(u))\,\dot\rho_\ep(u)
           +2\pa_i\,f(x^i_\ep(u))\,\dot x^i_\ep(u)\,\,\rho_\ep(u)\nn\\
     &=&\big(f(x^i_\ep(u))\,\rho_\ep(u)\big)\,\dot{\,}+\pa_i\,f(x^i_\ep(u))\,
        \dot x^i_\ep(u)\,\rho_\ep(u)\,\,.
\eea
The limit of the first term is easily seen to be $f(\aw)\,\dot\de(u)$, and
we are only left with the second term which contains the ``critical products''.
Inserting~(\ref{solreg}) and~(\ref{georeg}) respectively, we have
\bea \pa_i\,f(x^i_\ep)\,\dot x^i_\ep\,\rho_\ep&=&\pa_i\,f(x^i_\ep)\,
       (\dot x^i_0+\theta*\ddot x^i_\ep)\,\rho_\ep\nn\\
     &=&\pa_i\,f(x^i_\ep)\,\dot x^i_0\,\rho_\ep\,+\,\pa_i\,f(x^i_\ep)\,
       (\theta*[\frac{1}{2}\,\pa_i\,f(x^i_\ep)\,\rho_\ep]\,)\,\rho_\ep\,\,.
\eea      
The limit of the first term again is easily seen to give
$\dot x^i_0\,\pa_i\,f(\aw)\,\de(u)$, and we are finally left with the
task of computing the limit of the convolution term, i.e.,
the expression ``$\theta\de$''. 
Using 
\beq  (\theta*[\pa_i\,f(x^i_\ep)\,\rho_\ep]\,)\,(u)
      \,=\int\limits_{-\infty}^u\,\pa_i\,f(x^i_\ep(s))\,\rho_\ep(s)\,ds
      \,=\int\limits_{-1}^{u/\ep}\,\pa_i\,f(x^i_\ep(\ep s))\,\rho(s)\,ds
\eeq      
and again denoting by $\vphi$ a test function we get
\bea %&&\lim_{\ep\to 0}\Big(\frac{1}{2}\,\int\limits_{-\ep}^\ep\,
     %  \pa_if(x^i_\ep(u))\,
     %  [\,\int\limits_{-1}^{u/\ep}\,\pa_i\,f(x^i_\ep(\ep s))\,\rho(s)\,ds\,]\,
     %  \rho_\ep(u)\,\vphi(u)\,du\,\Big)\nn\\
    \lim_{\ep\to 0}&&\Big(\,
       \frac{1}{2}\int\limits_{-1}^1\pa_i\,f(x^i_\ep(\ep u))\,[\,\int\limits_
       {-1}^u\,\pa_i\,f(x^i_\ep(\ep s))\,\rho(s)\,ds\,]\,\rho(u)\,\vphi(\ep
       u)\,du\,\Big)\nn\\
    &&=\,\frac{1}{2}\left(\pa_i\,f(\aw)\,\right)^2\,\vphi(0)\,
       \underbrace{
       \int\limits_{-1}^1\,[\,\int\limits_{-1}^u\,\rho(s)\,ds\,]\,\rho(u)\,du
       }_{1/2}\nn\\
    &&=\,\frac{1}{4}\left(\pa_i\,f(\aw)\,\right)^2\,\vphi(0)\,\,.
\eea
Collecting things together we have
\beq  \lim_{\ep\to 0}\ddot v_\ep\,=\,
      f(\aw)\,\dot\de+\dot x^i_0\,\pa_i\,f(\aw)\,\de+
       \frac{1}{4} \left(\pa_i\,f(\aw)\,\right)^2\,\de
\eeq       
within distributions.
\vskip12pt

At the end of this section let us give a summary on what we have done so far.
We have regularized the distributionally ill-defined geodesic 
equations~(\ref{geo}) by replacing the $\de$-distribution by a (generic) class of 
mollifiers to obtain~(\ref{georeg}). These equations, 
although nonlinear, provide us with global solutions $v_\ep$ and $x^i_\ep$
for small $\ep$ (as shown in appendix~\ref{a1}), 
implicitly given by~(\ref{solreg}). Using the latter
formula we have shown that the smooth solutions have a regularization 
independent distributional limit given by
%Summing up we have shown that the distributional limit of the solutions to the 
%regularized geodesics equations~(\ref{georeg}) is given by
\vskip8pt
%\fbox{\parbox{16.3cm}{
\begin{eqnarray}\label{res1}
     v_\ep(u)&\to&v_0+\dot v_0\,(1+u)+f(0)\,\theta(u)+
        \pa_i\,f(0)\,(\dot x^i_0+\frac{1}{4}\pa_i\,f(0)\,)\,u_+\,\,,\nn\\
     x^i_\ep(u)&\to&x^i_0+\dot x^i_0\,(1+u)+\frac{1}{2}\,\pa_i\,
        f(0)\,u_+\,\,,     
\end{eqnarray}
 %} }
\vskip8pt
where we have used the abbreviation $f(0)=f(\aw)$.

Hence viewed distributionally the geodesics are given by refracted, broken
straight lines as expected. Of course equations~(\rf{res1}) coincide with 
the earlier results~\cite{fvp,ba}.
However from the point of view of our approach the (deeper) reason why 
here the ``rule'' $\theta\delta=(1/2)\,\de$ used by~\cite{fvp} 
(which in fact coincides with the ``determination of the point value'' 
$\theta(0)=1/2$ used by~\cite{ba}) leads to a physically reasonable result  
is the following: %twofold. First t
The geodesic equations involve {\em only one} singular object and 
hence the $\de$'s as well as the $\theta$'s appearing above share the same 
root: namely the $\de$-shaped wave profile. Hence, when regularizing the 
equations (which in fact corresponds to the physical idea of viewing the
impulsive wave as an idealized sandwich wave) both factors of the ill-defined 
product naturally involve the same regularization which immediately leads to 
the (regularization-independent) result $\rho_\ep\int\rho_\ep\to(1/2)\,\de$.
% which has to be regularized not only
%for technical reasons but also from physical grounds (bla bla...)
%Therefore the $\de$ as well as the $\theta$ should share the same

We thus conclude that the geodesic equation can be treated consistently by
regularization, leading to a regularization-independent distributional result.
This, of course, is only possible due to the relatively mild character of the 
singular terms which allows for a distributional limit of the solutions to the
regularized problem at all. 
However, we shall see in a moment that even in the considerably more
complicated case of the Jacobi equation our strategy can be applied
successfully.
%%%%%%%%%%%%%%%%%%%%%%%%%%%%%%%%%%%%%%%%%%%%%%%%%%%%%%%%%%%%%%%%%%%%%%%%%%%%%
\section{Jacobi Equation}\lb{j}
%%%%%%%%%%%%%%%%%%%%%%%%%%%%%%%%%%%%%%%%%%%%%%%%%%%%%%%%%%%%%%%%%%%%%%%%%%%%%
 In this section we solve the Jacobi equation for an impulsive pp-wave. To
keep formulas more transparent we restrict ourselves to the
axisymmetric case. % and treat only geodesics parallel to the axis. 
More precisely we restrict the function $f$ of the transverse
coordinates $x^i=(x,y)$ in the metric tensor~(\ref{metric}) to 
depend on the two-radius $r=\sqrt{x^2+y^2}$ only and 
% Then without loss of generality we shall 
work entirely within the $y=0$-hypersurface (initial conditions 
$x^2_0\equiv y_0=0=\dot x^2_0\equiv\dot y_0$).
Furthermore we take initial values $v_0=0=\dot x_0$.
With these assumptions equations~(\ref{res1}) simplify to
\bea v_\ep(u)&\to&\dot v_0\,(1+u)+f(x_0)\,\theta(u)+
                          \frac{1}{4}\,f'(x_0)^2\,u_+\,\,,\nn\\
     x^1_\ep(u)\equiv x_\ep(u)&\to&x_0+\fr{2}\,f'(x_0)\,u_+\,\,,\nn\\
     x^2_\ep(u)\equiv y_\ep(u)&=&0\,\,,
\eea     
where $f'$ denotes the derivative with respect to the single variable $r=x$.
Hence we shall deal with a geodesic tangent vector of the form
\beq T^a(u)=\left(\begin{array}{l}1\\
                           \dot v_0+f(x_0)\,\delta(u)+\fr{4}\,
                               f'(x_0)^2\,\theta(u)\\
                           \fr{2}\,f'(x_0)\,\theta(u)\\0
             \end{array}\right)\,\,,
\eeq             
where we are going to use the abbreviations
$A:=f(x_0)\,\delta(u)+(1/4)f'(x_0)^2\,\theta(u)$ and $B:=(1/2)f'(x_0)\,
\theta(u)$ for its components.%in the sequel.

Our next task is to compute the explicit form of the Jacobi equation 
$\frac{D^2N^a}{du^2}=-R\,^a_{bcd}T^b T^d N^c$ 
for a vector field $N^a(u)=(N^u(u),N^v(u),N^x(u),N^y(u))$ 
over the geodesic. After some (tedious) calculations we end up with the
following form of the system
\bea\label{jacobi}
     \ddot N^u &=&0\,\,,\nn\\
     \ddot N^v &=&2[N^x  f'\de]\,\dot{\,} -N^x  f'\dot
                     \de+[N^u  f\de]\,\ddot{\,}-
                     N^u  f''B^2 \de -
                     N^u  f'\dot B \de\,\,, \nn\\
     \ddot N^x &=&[\dot N^u f'+\fr{2}\,N^x  f'']\,\de +
                     \fr{2}\,f'N^u \dot\de\,\,, \nn\\
     \ddot N^y &=&0\,\,,
\eea
where we have suppressed the dependence on the parameter $u$ and the variable
$x$. Equations~(\ref{jacobi}) form a system of four coupled ODE's linear in the 
components of the vector field $N^a$ but nonlinear in the derivatives of the 
metric. From the fact that $B$ involves the step function we see immediately 
that (in the second equation) we again have to deal with distributionally 
ill-defined expressions, but now of even worse type than before. 
Indeed the term $\dot B\de$ is proportional to the ``square'' of the 
Dirac $\de$-distribution, and the term $B^2\de$ involves an expression 
``$\theta^2\de$''. Note, however, that the critical terms arise from the
second covariant derivative, where some of the Christoffel symbols get
multiplied, and not from the Riemann tensor which components are just
proportional to the $\de$-distribution.
To overcome these problems we
apply the same regularization procedure as in the case of the geodesic equation.
In particular, we use the regularized geodesic tangent vector 
\beq\lb{geot}
     T^a_\ep\,=\,\left(\begin{array}{l}  
                    1\\
                    \dot v_0+\theta*\ddot v_\ep\\
                    \theta*\ddot x_\ep\\
                    0
              \end{array}\right)\,%=:\,
%              left(\begin{array}{l}  
%                    1\\
%                    \dot v_0+A_\ep\\
%                    B_\ep\\
%                    0
%             \end{array}\right)\,
\,,
\eeq              
where $\ddot v_\ep(u)=f(x_\ep(u))\,\dot\rho_\ep(u)+2f'(x_\ep(u))\,\dot 
x_\ep(u)\,\rho_\ep(u)$,\quad $\ddot x_\ep(u)=(1/2)\,f'(x_\ep(u))\,\rho_\ep(u)$
and from now on we use the new abbreviations $A_\ep:=\theta*\ddot v_\ep$ and
$B_\ep:=\theta*\ddot x_\ep$. Denoting the regularized Jacobi field by
$N^a_\ep(u)=(N^u_\ep(u),N^v_\ep(u),N^x_\ep(u),N^y_\ep(u))$,
system~(\ref{jacobi}) takes the regularized form
\bea\label{rj}
     \ddot N^u_\ep &=&0\,\,,\nn\\
     \ddot N^v_\ep &=&2[N^x_\ep f'(x_\ep )\rho_\ep]\,\dot{\,}-
                        N^x_\ep f'(x_\ep )\dot\rho_\ep +
                        [N^u_\ep f(x_\ep )\rho_\ep]\,\ddot{\,}-\nn\\
                   &&N^u_\ep f''(x_\ep )[\theta*\ddot
                        x_\ep]^2 \rho_\ep -
                        N^u_\ep f'(x_\ep )\ddot x_\ep \rho_\ep\,\,, \nn\\
     \ddot N^x_\ep &=&[\dot {N^u_\ep} f'(x_\ep )+
                         \fr{2}\,N^x_\ep f''(x_\ep )]\rho_\ep 
                         +\fr{2}\,f'(x_\ep )N^u_\ep \dot\rho_\ep \,\,,\nn\\
     \ddot N^y_\ep &=&0\,.
\eea
                        
To maintain transparency of formulae we choose appropriate and simple 
initial conditions  on the Jacobi field $N^a_\ep$ at
$u=-1$, i.e., 
\bea \label{icrj} N^a_\ep(-1)&=&(0,0,0,0)\,\,,\nn\\
     \dot N^a_\ep(-1)&=&(a,b,0,0)\,\,,
\eea
for all $\ep$. Note that this corresponds to a focal point at $u=-1$, and that
``nearby'' geodesics have relative initial velocities only in the $u$- and 
$v$-direction.
%(??????????focal point??????????????)

 The first and the last equation of system~(\ref{rj}) are easily solved
to give 
\vskip8pt
%\fbox{\parbox{16.3cm}{
\bea\lb{sl} N^u_\ep(u)&=&a(1+u)\,\,,\nn\\
          N^y_\ep(u)&=&0\,.
\eea
%}}
\vskip20pt

For the remaining, more complicated equations of system~(\ref{rj})
we apply the same strategy as in the preceeding section.
Since the equations are linear in the components of the deviation vector field
we % The right hand sides
%of the regularized equations again are smooth and compactly supported, hence
%we 
obtain globally defined (smooth) solutions which, due 
to the initial conditions~(\ref{icrj}), are implicitly given by 
\bea\lb{ins}\lb{bbl} 
   N^v_\ep(u)&=&b\,(1+u)+[\theta*\theta*\ddot N^v_\ep]\,(u)\,\,,\nn\\
   N^x_\ep(u)&=&[\theta*\theta*\ddot N^x_\ep]\,(u)\,\,.
\eea
Again it suffices to compute the distributional limits of the right hand 
sides of equation~(\ref{rj}), since by continuity of the convolution we 
immediately get the limits of $N^v_\ep$ and $N^x_\ep$. 
\vskip12pt

We start with the third equation of the system~(\ref{rj}). 
The main problem we have to face here is due to the fact that the unknown 
function $N^x_\ep$ (which, in the limit, we cannot even expect to be 
continuous at $u=0$) appears on the right hand side.
Inserting the initial conditions~(\ref{icrj}) and the solutions~(\ref{sl}) 
we get 
\beq \label{eq3}\ddot
N^x_\ep(u)\,=\underbrace{\,af'(x_\ep(u))\,\rho_\ep(u)}_{I_\ep}
            +\underbrace{\fr{2}\,f'(x_\ep(u))\,a(1+u)\dot\rho_\ep(u)}
                       _{II_\ep}
            +\underbrace{\fr{2}\,N^x_\ep(u)f''(x_\ep(u))\,\rho_\ep(u)}
                       _{III_\ep}\,.
\eeq
%To keep formulas transparent we split up the individual terms in
%equation~(\ref{eq3}) according to their different $u$-dependences, using the 
%shorthand notation
%\beq\label{eq28}
%\ddot N^x_\ep(u)\,
%    =\,I_\ep(u)\rho_\ep(u)+II_\ep(u)\dot\rho_\ep(u)+
%                           II_\ep(u)u\dot\rho_\ep(u)\,,
%\eeq
%with $I_\ep(u):=af'(x_\ep(u))+(1/2)\,N^x_\ep(u)f''(x_\ep(u))$ and
%$II_\ep(u):=(1/2)\,af'(x_\ep(u))$.
The distributional limits of the first two terms in equation~(\rf{eq3})
are easily seen (using similar methods as in the previous section) to be
\bea\lb{29}
  \lim_{\ep\to 0}I_\ep&=&%_\ep\dot\rho_\ep&=
                             af'(x_0)\,\de\,\,,\nn\\            
  \lim_{\ep\to 0}II_\ep&=&\frac{1}{2}\,af'(x_0)\,(\dot\delta-\de)
              -\frac{1}{8}\,af'(x_0)f''(x_0)\,\de\,.
%  \lim_{\ep\to 0}II_\ep u\dot\rho_\ep&=
%            &-\frac{1}{2}\,af'(x_0)\,\de\,.
\eea
However, the last term in equation~(\ref{eq3})
%\beq 
%    III_\ep(u)\rho_\ep(u)\,=\,\frac{1}{2}\,N^x_\ep(u)f''(x_\ep(u))\,\rho_\ep(u)
%\eeq
requires a more detailed analysis.
%, since the term labeled by $IV_\ep$ involves
%the unknown function $N^x_\ep$. First we note that the ``easy'' term
%$III_\ep\rho_\ep$ has the distributional limit
%\beq\lb{31} \lim_{\ep\to0}III_\ep\,\rho_\ep  =  af'(x_0)\,\de\,.\eeq
%Next we calculate the limit of
%\beq IV_\ep\rho_\ep\,=\,\frac{1}{2}\,N^x_\ep f''(x_\ep)\rho_\ep
%     =\frac{1}{2}\,f''(x_\ep)\,\rho_\ep\,\left[\theta*\theta*(I_\ep\rho_\ep+
%       II_\ep\dot\rho_\ep+II_\ep\,u\dot\rho_\ep)\,\right]\,,
%\eeq
%where we have used equations~(\ref{ins}) and~(\ref{eq28}).
Inserting again $\theta*\theta*\ddot N^x_\ep$ for the unknown function $N^x_\ep$
and denoting by $\vphi$ a test function we obtain
\bea %&\,&\frac{1}{2}\int\limits_{-\ep}^{\ep}\vphi(u)f''(x_\ep(u))\,\rho_\ep(u)
     %  \int\limits_{-\ep}^u\int\limits_{-\ep}^s 
     %  [I_\ep(r)\rho_\ep(r)+II_\ep(r)\dot\rho_\ep(r)
     %                      +II_\ep(r)r\dot\rho_\ep(r)]\,dr\,ds\,du
     %\\ &=&
     &&III_\ep(u)=\nn\\
     &&\frac{1}{2}\,\ep\int\limits_{-1}^1\vphi(\ep u)f''(x_\ep(\ep u))\rho(u)
      \int\limits_{-1}^u\int\limits_{-1}^s[ a f'(x_\ep(\ep r))
      +\fr{2}\,N^x_\ep(\ep r)f''(x_\ep(\ep r))]\,\rho(r)\,dr\,ds\,du
     \nn\\&& +
     \frac{1}{4}\,a\int\limits_{-1}^1\vphi(\ep u)f''(x_\ep(\ep u))\rho(u)\int
      \limits_{-1}^u\int\limits_{-1}^s f'(x_\ep(\ep r))\,\dot\rho(r)\,
      dr\,ds\,du
     \nn\\ \lb{eq36}&&+
     \frac{1}{4}\,a\ep\int\limits_{-1}^1\vphi(\ep u)f''(x_\ep(\ep u))\rho(u)
       \int\limits_{-1}^u\int\limits_{-1}^s f'(x_\ep(\ep r))\, r
       \dot\rho(r)\,dr\,ds\,du\,\,.
\eea       
The last term in equation~(\ref{eq36}) goes to zero, since $\vphi,\,f',\,f''$ 
are smooth functions and $x_\ep$ is bounded uniformly on compact sets 
(see appendix~\ref{a1}).
%(cf. equation~(\ref{eq11})\,) 
%and the integration is to be taken over a compact set only. 
For the same reasons the second summand in equation~(\ref{eq36}) approaches 
the limit
\beq\lb{35}\frac{1}{8}\,a\,\vphi(0)\,f'(x_0)\,f''(x_0)\,,\eeq
whereas the first line also vanishes in the limit, since $N^x_\ep(\ep r)$ is 
bounded on the compact range of integration. (Note, however, that 
$\lim\,N^x_\ep$ is not continuous at $0$!)
In some more detail the latter argument can be seen from
\bea N^x_\ep(\ep u)%&=&\int\limits_{-\ep}^{\ep u}\int\limits_{-\ep}^s
                     %\left\{[af'(x_\ep(r))+\fr{2}\,N^x_\ep(r)f''(x_\ep(r))]\,
                     %\rho_\ep(r)
                     %+\fr{2}\,a f'(x_\ep(r))\,
                     %      [\dot\rho_\ep(r)+r\dot\rho_\ep(r)]
                     %\right\}\,dr\,ds\nn\\
                   &=&\int\limits_{-1}^u\int\limits_{-1}^s\left\{af'(x_\ep(\ep
                      r))\ep\rho(r)+\fr{2}\,af'(x_\ep(\ep r))[\dot\rho(r)+
                      r\ep\dot\rho(r)]\right\}\,dr\,ds\nn\\
                   &&+\frac{1}{2}\int\limits_{-\ep}^{\ep u}\int\limits_{-1}
                      ^{s/\ep}N^x_\ep(\ep r)f''(x_\ep(\ep r))\rho(r)dr\,ds\,,
\eea                      
where we have now used equation~(\ref{ins}) for the third time. 
Hence we have for $u\leq T$, where $T$ is some constant, and for fixed $\ep$
\bea \mid N^x_\ep(\ep u)\mid
    &\leq&C+C'\int\limits_{-\ep}^{\ep u}\int\limits_{-1}^{s/\ep}
           \mid N^x_\ep(\ep r)\mid\,\mid\rho(r)\mid\,dr\,ds\nn\\
    %&=&
    %  C+C'\int\limits_{-1}^{u}\int\limits_{\ep r}^{\ep u}
    %       \mid N^x_\ep(\ep r)\mid\,\mid\rho(r)\mid\,ds\,dr\nn\\
    %&\leq&C+C'\int\limits_{-1}^u\int\limits_{-1}^1
    %       \mid N^x_\ep(\ep r)\mid\,\mid\rho(r)\mid\,ds\,dr
    %       \mbox{\quad ($\ep$ small)}\,\leq\,\nn\\
    %&\leq&
    %\nn\\
    &\leq&C+2C'\int\limits_{-1}^{u}\mid N^x_\ep(\ep r)\mid\,\mid\rho(r)\mid\,dr
    \,,
\eea
where $C=C(T)$ and $C'$ denote constants. From Gronwall's inequality we gain
\beq \mid N^x_\ep(\ep u)\mid\,\leq\,C\,
         e^{2C'\int_{-1}^u\mid\rho(r)\mid\,dr}\,\,,
%                            &\leq& constant\,.
\eeq                            
which is bounded by a constant for $u\leq T$.

Collecting things together (equations~(\ref{29}) and~(\ref{35})\,)
we find that within distributions
\beq \ddot N^x_\ep\,\to\,\frac{1}{2}\,a\,f'(x_0)\,(\de+\dot\de)\,,\eeq

which gives us imediately the distributional limit of the regularized solution 
of the third equation of system~(\ref{rj}), i.e.,
\vskip8pt
%\fbox{\parbox{16.3cm}{
\beq N^x_\ep(u)\,\to\,\frac{1}{2}\,a f'(x_0)\,(u_++\theta(u))\,.\eeq
%}}
\vskip20pt

Now we face the most complicated equation of system~(\ref{rj}) which contains
the distributionally ill-defined expressions, namely
\bea\lb{leq}
    \ddot N^v_\ep(u)
    &=&2[N^x_\ep(u)\,f'(x_\ep(u))\,\rho_\ep(u)\,]\,\dot{\,}
     +[N^u_\ep(u)\,f(x_\ep(u))\,\rho_\ep(u)\,]\,\ddot{\,}\nn\\
    &-&N^u_\ep(u)\,f''(x_\ep(u))\,[\theta*\ddot x_\ep]^2(u)\,\rho_\ep(u)%\nn\\
    -N^x_\ep(u)\,f'(x_\ep(u))\,\dot\rho_\ep(u)\nn\\
    &-&N^u_\ep(u)\,f'(x_\ep(u))\,\ddot x_\ep(u)\,\rho_\ep(u)\,,
\eea
where we have inserted all the dependences explicitly. Note that in addition to
the obvious troublesome fifth term  which involves the ``square'' of the 
$\de$-distribution ($\ddot x_\ep\propto\rho_\ep$ !), 
also the fourth term contains
divergent and regularization-dependent contributions, since the derivatives 
of $f(x)$ and $N^x_\ep$ produce $\de$- and step functions respectively. 
We are going to calculate the distributional limits of the right hand side of
equation~(\ref{leq}) term by term, which can easily be done for the first
three ones using yet well known techniques.
%we obtain within distributions\vskip12pt

\noindent
\underline{First, second and third term:}
\bea \lb{(i)}
 \lim_{\ep\to 0}\,2[N^x_\ep(u)\,f'(x_\ep(u))\,\rho_\ep(u)\,]\,\dot{\,}
    &=&\fr{2}\,a\,f'(x_0)^2\,\dot\de\,\,,\nn\\
\lb{(ii)}
 \lim_{\ep\to 0}\,\,[N^u_\ep(u)\,f(x_\ep(u))\,\rho_\ep(u)\,]\,\ddot{\,}\,
          &=&a\,f(x_0)\,\ddot\de\,\,,\nn\\
\lb{(iii)}
 \lim_{\ep\to 0}\,N^u_\ep(u)\,f''(x_\ep(u))\,[\theta*\ddot x_\ep]^2(u)\,
               \rho_\ep(u)\,
    &=&\fr{12}\,a\,f'(x_0)^2\,f''(x_0)\,\de
\eea
\vskip12pt

We now come to the expressions involving the divergent and
regularization-dependent factors. 
We outline the calculations in some detail.

\noindent
\underline{Fourth term:} 
   $(iv):=-N^x_\ep(u)\,f'(x_\ep(u))\,\dot\rho_\ep(u)$\newline
By integration by parts we obtain ($\vphi$ again a test function)
\bea
  &-&\il_{-\ep}^\ep\vphi(u)\,N^x_\ep(u)\,f'(x_\ep(u))\,\dot\rho_\ep(u)\,du
 =\underbrace{
    \il_{-1}^1\dot\vphi(\ep u)\,N^x_\ep(\ep u)\,f'(x_\ep(\ep u))\,\rho(u)\,du 
    }_{A} \\
 &&+\underbrace{\il_{-1}^1\vphi
          (\ep u)\,N^x_\ep(\ep u)\,f''(x_\ep(\ep
     u))\,\dot x_\ep(\ep u)\,\rho(u)\,du}_{B}+
     \underbrace{\il_{-1}^1 \vphi(\ep u)\dot N^x_\ep(\ep
     u)\,f'(x_\ep(\ep u))\rho(u)\,du}_{C}.\nn
\eea     
Inserting $\theta*\theta*\ddot N^x_\ep$ according to
equations~(\ref{ins}) and~(\ref{eq3}) for $N^x_\ep$ %For $A$ this gives
we find for the expressions labelled $A$ and $B$
\bea\lb{46} A&\to&-\fr{4}\,a\,f'(x_0)^2\,\dot\de\,\,,\\
\lb{48} B&\to&\fr{12}\,a\,f'(x_0)^2\,f''(x_0)\,\de\,.\eea

Finally we have for the most complicated expression
\bea\label{abuse} C
    &=&\underbrace{
       \frac{1}{2\ep}\,a\il_{-1}^1\vphi(\ep u)\,f'(x_\ep(\ep u))\,\rho(u)
       \il_{-1}^uf'(x_\ep(\ep s))\,\dot\rho(s)\,ds\,du}_{C1}\nn\\
    &\,&+\underbrace{
       \frac{1}{2}\,a\il_{-1}^1\vphi(\ep u)\,f'(x_\ep(\ep u))\,\rho(u)
       \il_{-1}^uf'(x_\ep(\ep s))\,s\dot\rho(s)\,ds\,du}_{C2}\\
    &\,&+\underbrace{
       \il_{-1}^1\vphi(\ep u)\,f'(x_\ep(\ep u))\,\rho(u)
       \il_{-1}^u\big[a\,f'(x_\ep(\ep s))+\fr{2}N^x_\ep(\ep s)\,f''(x_\ep(\ep
       s))\,\big]\,\rho(s)\,ds\,du}_{C3}\,.\nn
\eea       
The term labeled $C1$ diverges due to the factor $\ep^{-1}$.
%, since the integrand is bounded. an we have a
%factor $\ep^{-1}$ outside the integral. 
However, as we shall see later, this 
expression combined with another one arising from the fifth term of
equation~(\ref{leq}) will provide us with a finite result.

The term labelled $C2$ involves a regularization-dependent factor, i.e.,
\beq\lb{50}
    C2\,\to\,\frac{1}{2}\,af'(x_0)^2\,\de\,\il_{-1}^1\rho(u)
           \il_{-1}^us\dot\rho(s)\,ds\,dt \,.
\eeq     
As in the case of the previous term $C1$, this problem will be resolved 
later on, so that we are now left with the task of computing the limit 
of $C3$. Again using similar techniques we obtain
\beq \lb{45}  C3\,\to\,
  \frac{1}{2}\,af'(x_0)^2\,\de+\frac{1}{24}\,af'(x_0)^2\,f''(x_0)\,\de\,.
\eeq
Collecting together the results of equations~(\ref{46}),~(\ref{48}) 
and~(\rf{45}), we finally have
\beq\lb{(iv)} (iv)-C1-C2\,\to\,-\fr{4}\,af'(x_0)^2\,\dot\de
                  +\fr{8}\,af'(x_0)^2\,f''(x_0)\,\de
                  +\fr{2}\,af'(x_0)^2\,\de\,,
\eeq                  
where now, by an abuse of notation $C1$ and $C2$ denote the respective 
quantities in equation~(\ref{abuse}) without integration over the test 
function.
\vskip12pt

Now we finally come to the last term in equation~(\ref{leq}) which will resolve
the problems with the expressions $C1$ and $C2$.
% contained in the second term.
\newline
\underline{Fifth term:} $(v):=-N^u_\ep(u)\,f'(x_\ep(u))\,\ddot x_\ep(u)\,
\rho_\ep(u)$\newline
Inserting solution~(\rf{sl}) we find
\bea
    -\il_{-\ep}^\ep\vphi(u)\,N^u_\ep(u)\,f'(x_\ep(u))\,
       \ddot x_\ep(u)\,\rho_\ep(u)\,du
    \nn\\
    =\underbrace{-\fr{2\ep}\,a\il_{-1}^1\vphi(\ep u)\,f'(x_\ep(\ep
       u))^2\,\rho(u)^2\,du}_{(*)}
    &\,&\underbrace{-\fr{2}\,a\il_{-1}^1\vphi(\ep u)\,f'(x_\ep(\ep
       u))^2\,u\rho(u)^2\,du}_{(**)}\,.
\eea       
We see now that the expression $(*)$ diverges,
% since the integrand again is bounded,
but that the sum $C1+(*)$ converges. More precisely, by integration by parts 
we get
\bea C1+(*)%&=&\fr{2\ep}\,a\il_{-1}^1\vphi(\ep u)\,f'(x_\ep(\ep
            %  u))\rho(u)\big[
            % \il_{-1}^uf'(x_\ep(\ep s))\,\dot\rho(s)\,ds
            % \,-\,f'(x_\ep(\ep u))\,\rho(u)\,\big]\,du\nn\\
           &=&-\fr{4}a\il_{-1}^1\vphi(\ep u)f'(x_\ep(\ep u))\rho(u)
             \il_{-1}^uf''(x_\ep(\ep s))\rho(s)
             \il_{-1}^sf'(x_\ep(\ep r))\rho(r)drdsdu\nn\\
           &\to&-\fr{4}\,a\,\vphi(0)\,f'(x_0)^2\,f''(x_0)
             \underbrace{
             \il_{-1}^1\rho(u)\il_{-1}^u\rho(s)\il_{-1}^s\rho(r)\,dr\,ds\,du}
             _{1/6}\nn\\
           &=&-\fr{24}\,a\,\vphi(0)\,f'(x_0)^2\,f''(x_0)\,.
\eea          
Hence we have the distributional limit
\beq \lb{C1} C1+(*)\,\to\,-\fr{24}\,a\,f'(x_0)^2\,f''(x_0)\,\de\,.\eeq

On the other hand expression $(**)$ contains a regularization-dependent 
factor which is proportional to the one contained in the term $C2$.
%In some more detail we have in the sense of distributions
More precisely, from equation~(\rf{50}) and the fact that
\beq (**)\,\to\,-\fr{2}\,a\,f'(x_0)^2\,\de\,\il_{-1}^1u\rho(u)^2\,du\eeq
%and
%\beq C2\,\to\,\fr{2}\,a\,f'(x_0)^2\,\de\,\il_{-1}^1\rho(u)\il_{-1}^u
%     s\dot\rho(s)\,ds\,du\,.
%\eeq
we have
\bea\lb{C2} C2+(**)&\to&\fr{2}\,a\,f'(x_0)^2\,\de\,
     \underbrace{\big[\il_{-1}^1\rho(u)\,[\il_{-1}^u
     s\dot\rho(s)\,ds-u\rho(u)]\,du\big]}_{1/2}\nn\\
     &=&-\frac{1}{4}\,a f'(x_0)^2\,\de\,.
\eea     
%hence
%\beq \lb{C2}C2+(**)\,\to\,-\fr{4}\,a\,f'(x_0)^2\,\de\,.\eeq
%\vskip12pt

We are now ready to collect together all five terms of equation~(\ref{leq}),
i.e., of the second equation of the regularized system~(\ref{rj}).
From equations~(\ref{(iii)}), (\ref{(iv)}), 
(\ref{C1}) and~(\rf{C2}) we get
the following result:
\beq 
 \ddot N^v_\ep\,\to\,a\,f(x_0)\,\ddot\de+\fr{4}\,a\,f'(x_0)^2\,(\dot\de+\de)\,,
\eeq
which, using initial condition~(\rf{icrj}), we readily integrate to 
\vskip8pt
%\{\parbox{16.3cm}{
\beq
 N^v_\ep(u)\,\to\,b\,(1+u)+a\,\big[f(x_0)\,\de+\fr{4}f'(x_0)^2\, 
 (\theta(u)+u_+)\,\big]\,.
\eeq
%}}
\vskip20pt

Finally we have for the whole system~(\ref{rj}) the following (distributional
limits of the) solutions
\vskip8pt
%\fbox{\parbox{16.3cm}{
\bea
  N^u_\ep(u)&=&a\,(1+u)\,\,,\nn\\
  N^v_\ep(u)&\to&b\,(1+u)+a\,\big[f(x_0)\,\de+\fr{4}f'(x_0)^2\,
                 (\theta(u)+u_+)\,\big]\,\,,\nn\\
  N^x_\ep(u)&\to&\fr{2}\,a\,f'(x_0)\,(\theta(u)+u_+)\,\,,\nn\\
  N^y_\ep(u)&=&0\,.
\eea               
%}}
\vskip12pt

Hence, viewed distributionally, the Jacobi field suffers a kink, a jump and a
$\de$-like pulse in the $v$-direction as well as a kink and jump in the 
$x$-direction overlapping the linear flat space behavior. These effects can be
understood heuristically from the corresponding behavior of the geodesics, given
by equation~(\ref{res1}). The constant factor $a$, which gives the
``scale'' of all the nonlinear effects, arises from the ``time advance'' 
of the ``nearby'' geodesics, represented by the initial velocity of the Jacobi
field in the $u$-direction (cf. initial conditions~(\ref{icrj})\,). 

Note, however, that this ``time advance'' is not the only effect generically
generating deviations from the flat space behavior, but rather arises as an 
artifact of our initial conditions. %(recall that we {\em only} treated 
%{\em axiparallel} ``nearby'' geodesics). 
One can easily show that
different initial conditions on the deviation vector field, even without ``time
advance'', produce kinks and jumps as well. For example, let
\beq N^a_\ep(-1)\,=\,(0,0,0,0)\,
                \mbox{ and }\, \dot N^a_\ep(-1)\,=\,(0,a,b,0)\,;\eeq
then a similar (but now even simpler) calculation leads to the following 
(distributional limits of the) regularized Jacobi field
\bea\lb{ssol}
   N^u_\ep&=&0\,\,,\nn\\
   N^v_\ep&\to&a(1+u)+\fr{4}\,b\,f'(x_0)\,f''(x_0)\,u_+
                     +b\,f'(x_0)\,\theta(u)\,\,,\nn\\
   N^x_\ep&\to&b(1+u)+\fr{2}\,b\,f''(x_0)\,u_+\,\,,\nn\\
   N^y_\ep&=&0\,\,.
\eea   
The kink of the $x$-component of the deviation field now arises from the fact
that a ``nearby'' geodesic passes the shock at an $x$-value of $x_0+b$, hence
according to equation~(\ref{res1}) suffers a kink of ``strength'' $(1/2)\,
f'(x_0+b)$. Taylor expansion yields $f'(x_0+b)\approx f'(x_0)+f''(x_0)\,b$, 
such that the kink difference of ``nearby'' geodesics is given by $(1/2)\,
f''(x_0)\,b$, which is exactly the factor given in the third equation
of~(\ref{ssol}). The kink and jump in $v$-direction can be explained by similar
heuristic arguments.
%%%%%%%%%%%%%%%%%%%%%%%%%%%%%%%%%%%%%%%%%%%%%%%%%%%%%%%%%%%%%%%%%%%%%%%%%%%% 
\section{Conclusion}\lb{c}

We have shown that the geodesic and Jacobi equation of impulsive pp-waves can 
be treated consistently using the distributional form of the space time
metric~(\ref{metric}). 

The main problem was to deal with products ill-defined
within Schwartz's linear distribution theory. The geodesic equation involves
terms proportional to ``$\theta\de$'' while the Jacobi equation, although 
linear in the components of the deviation vector field, contains even 
more singular terms like ``$\de^2$'' and ``$\theta^2\de$''. 
Our strategy consisted essentially in a
careful and general regularization procedure, replacing the
$\de$-distribution by a mollifying sequence $\rho_\ep$. While mathematically 
sound, this procedure corresponds to the physical idea of viewing the 
impulsive wave as the limiting case of a sandwich wave of ever decreasing 
support but constant strength. 
Our approach leads to smooth solutions of the regularized 
equations which possess a regularization-independent limit within the 
vector space of distributions. However, this is not obvious
% matter of course
(cf. for example~\cite{stein}), but rather should strengthen our trust in 
impulsive waves as reasonable solutions of Einstein's equations. 
Furthermore our distributional ``solutions'' of the geodesic and Jacobi equation
perfectly coincide with physical expectations, showing that the geometry 
of impulsive pp-waves can be described in a consistent way using the 
distributional form of the metric, as long as one applies proper regularization
methods as opposed to ``ad-hoc strategies'' involving certain ``multiplication
rules''. In fact when dealing with nonlinear operations (of 
a certain complexity) on singular, i.e., distributional data, reliable 
results can only be achieved by careful regularization procedures 
(cf. the comments on ``common errors'' even in the mathematical literature 
given by H\'ajek~\cite{hajek}).

However from an even more formal point of view, the approach taken in this
paper suffers from the fact that we do not really have an elaborated
solution concept for the regularized equations. Such a notion is provided
by the theory of Nonlinear Generalized Functions~\cite{mo,co1,co2} 
due to J.\,F. Colombeau and others, where - loosely speaking - the 
regularized sequence is viewed as {\em one} mathematical object, i.e., a
generalized function. Recently, singular, nonlinear ODEs (exactly the type
of equations we were dealing with) have been studied in the context of 
this new mathematical framework~\cite{moODE,kunzDISS}, providing a powerful
tool to handle both, {\em singular} data and especially this type of 
{\em nonlinear} differential equations. Future work will be concerned
with the analysis of the geodesic and Jacobi equation for pp-waves from
the point of view of this new solution concept.
%%%%%%%%%%%%%%%%%%%%%%%%%%%%%%%%%%%%%%%%%%%%%%%%%%%%%%%%%%%%%%%%%%%%%%%%%%%%%
%%%%%%%%%%%%%%%%%%%%%%%%%%%%%%%%%%%%%%%%%%%%%%%%%%%%%%%%%%%%%%%%%%%%%%%%%%%%%
\subsection*{Acknowledgement}
The author owes special thanks to M. Kunzinger for great support with the
calculations and M. Oberguggenberger and  M. Grosser for many helpful
discussions.

This work was supported by Austrian Academy of Science, Ph.D. programme, 
Grant \#338.
%%%%%%%%%%%%%%%%%%%%%%%%%%%%%%%%%%%%%%%%%%%%%%%%%%%%%%%%%%%%%%%%%%%%%%%%%%%%%
%%%%%%%%%%%%%%%%%%%%%%%%%%%%%%%%%%%%%%%%%%%%%%%%%%%%%%%%%%%%%%%%%%%%%%%%%%%%%
%APPENDICES
\begin{appendix}
\section*{}\lb{a1}

In this appendix we prove (for small $\ep$) the global existence 
and uniform boundedness (in $\ep$) of the solutions to the last equation of 
system~(\ref{georeg}). For simplicity we only treat the one-dimensional case 
and write $g$ instead of $Df$. The main idea is to use $\|\,\|_1$-bounds on 
the regularizing sequence $\rho_\ep$. We start with the following statement.
\vskip12pt
{\bf Proposition: }{\it 
Let $g:\R\to\R$ smooth and $\rho\in\DD([-1,1]),\,\int\rho=1$, $\rho_\ep(x)
:=(1/\ep)\,\rho(x/\ep)$. For fixed $\ep$ consider the differential equation
\beq \label{dgl} \ddot x_\ep(t)\,=\,g(x_\ep(t))\,\rho_\ep(t)\,,\eeq
with initial conditions $x_\ep(-1)=x_0$ and $\dot x_\ep(-1)=\dot x_0\,>
0\,.$
Then there exists a unique solution $x_\ep(t)$ on the interval 
$J_\ep:=[-1,-\ep+\al]$, where
\beq\al:=min\left\{\frac{b}{\|g\|_{\infty,I}\,\|\rho\|_1+ \dot x_0},
\frac{1}{2L}\,\right\}\qquad (b> 0)\,\,,\eeq
$I:=\{x\in\R:\,\mid x-x_0\mid\leq b+\dot x_0\,\}$ and
$L$ is a Lipschitz constant for $g$ on $I$.}
\vskip12pt

The {\it Proof }is just a modification of the proof of the classical 
(first order) ODE existence and uniqueness theorem. We 
work on the (nonempty, closed) subset $X_\ep:=\{x_\ep\in C(J_\ep):\,
\mid x_\ep(t)-x_0\mid\leq b+\dot x_0\,\}$ of the Banach space of 
continuous functions on the interval $J_\ep$. Since no first order
derivative of $x_\ep$ enters the right hand side of equation~(\ref{dgl})
we may define the integration operator $A$ by
\beq [A\,x_\ep]\,(t)\,:=\,x_0+\dot x_0(t+1)+\il_{-\ep}^t\il_{-\ep}^s
 \,g(x_\ep(r))\,\rho_\ep(r)\,dr\,ds\,\,.\eeq
By construction $Ax_\ep\in C(J_\ep)$, but we even have $Ax_\ep\in X_\ep$ since 
\bea\mid\,[A\,x_\ep]\,(t)-x_0\,\mid&\leq&\dot x_0(t+1)+
\il_{-\ep}^t\il_{-\ep}^s \,\mid g(x_\ep(r))\mid\,\mid\rho_\ep(r)\mid\,dr\,ds
\nn\\
&\leq&\al(\dot x_0+\|g\|_{\infty,I}\|\rho\|_1)+\dot x_0\,\leq\,b+\dot x_0
\eea
By a similar estimate one shows that $A$, in fact, is a contraction on
$X_\ep$. Hence by the fixed point theorem we have a unique solution $x_\ep\in
X_\ep\,.\quad\Box$
\vskip12pt

Since $\al$ is independent of $\ep$ we have for small $\ep$:
$-\ep+\al\ge\ep$. But at $t=\ep$ the right hand side of the differential
equation~(\ref{dgl}) has already been ``turned off''. 
Hence for small $\ep$ the solutions are defined on the whole real line.

Furthermore, since for all $\ep$ the solution lies in $X_\ep$, $x_\ep$ is
bounded uniformly in $\ep$ for say $t\leq\al/2$ (for $\ep$ small enough). 
For larger values of $t$ $x_\ep$ grows only linearly with $\dot x_\ep$ 
which is bounded by $\dot x_0+\|g\|_{\infty,I}\|\rho\|_1$. 
Hence (for small $\ep$) $x_\ep(t)$ is bounded uniformly in $\ep$ on every 
compact set.

\end{appendix}
%%%%%%%%%%%%%%%%%%%%%%%%%%%%%%%%%%%%%%%%%%%%%%%%%%%%%%%%%%%%%%%%%%%%%%%%%%%%%
%%%%%%%%%%%%%%%%%%%%%%%%%%%%%%%%%%%%%%%%%%%%%%%%%%%%%%%%%%%%%%%%%%%%%%%%%%%%%
%literature
                 
%%%%%%%%%%%%%%%%%%%%%%%%%%%%%%%%%%%%%%%%%%%%%%%%%%%%%%%%%%%%%%%%%%%%%%%%%%
\end{document}